\documentclass[11pt]{article}
\usepackage{a4wide}
\usepackage{epsf}
\usepackage{color}
\usepackage{graphicx}
\usepackage{amsmath}
\usepackage{amssymb}
\usepackage{latexsym}

\DeclareMathOperator{\diag}{diag}

\newcommand{\V}{\mathcal{V}}
\newcommand{\lamhat}{\hat\lambda}

\newcommand{\mKK}{m_{\text{KK}}}

\begin{document}
\title{Diagonal Kaluza-Klein expansion under brane localized potential \bigskip}
\author{
	Naoyuki Haba\thanks{E-mail: \tt haba@phys.sci.osaka-u.ac.jp}, 
	Kin-ya Oda\thanks{E-mail: \tt odakin@phys.sci.osaka-u.ac.jp}, and 
	Ryo Takahashi$^{a,b,}$\thanks{E-mail: \tt Ryo.Takahashi@mpi-hd.mpg.de}\\
	\\
	{\it\normalsize Department of Physics, Osaka University,}\\{\it\normalsize Osaka 560-0043, Japan}\\ $^a${\it\normalsize Yukawa Institute for Theoretical Physics, Kyoto University,}\\{\it\normalsize Kyoto 
606-8502, Japan}\\$^b${\it\normalsize Max-Planck-Institute f$\ddot{u}$r Kernphysik, Postfach 10 39 80, 69029}\\
{\it\normalsize Heidelberg, Germany}\\
	\bigskip\\}
\maketitle
\begin{abstract}\noindent 
We clarify and study our previous observation that, under a compactification with boundaries or orbifolding, vacuum expectation value of a bulk scalar field can have different extra-dimensional wave-function profile from that of the lowest Kaluza-Klein mode of its quantum fluctuation, under presence of boundary-localized potentials which would be necessarily generated through renormalization group running.
For concreteness, we analyze the Universal Extra Dimension model compactified on orbifold $S^1/Z_2$, with brane-localized Higgs potentials at the orbifold fixed points.
We compute the Kaluza-Klein expansion of the Higgs and gauge bosons in an $R_\xi$-like gauge by treating the brane-localized potential as a small perturbation.
We also check that the $\rho$ parameter is not altered by the brane localized potential.
  \end{abstract}
\vfill
\hfill OU-HET-643
\newpage

\section{Introduction}
The five dimensional Quantum Field Theory (QFT), compactified on the orbifold $S^1/Z_2$, has been paid much attention as the basis for 
	the extra dimensional standard model with bulk gauge bosons~\cite{Nath:1999fs,Masip:1999mk,Rizzo:1999br,Strumia:1999jm,Carone:1999nz},
	Universal Extra Dimension (UED) model~\cite{Appelquist:2000nn,Appelquist:2002wb}, 
	Higgsless model~\cite{Csaki:2003dt},
	gauge-Higgs unification models (see e.g.~\cite{Serone:2005ds} and references therein), and also
	the supergravity models~\cite{Kugo:2000hn,Zucker:2000ks,Kugo:2000af,Fujita:2001bd}.
The five dimensional QFT on $S^1/Z_2$ is also the starting point for the QFT in the warped space,\footnote{
	Originally Randall and Sundrum proposed it without any bulk field other than graviton~\cite{Randall:1999ee}. See also~\cite{Oda:2008tk} for a possible regularization of the negative tension brane.
	}
	which is again utilized in the warped version of the bulk standard~\cite{Agashe:2003zs,Albrecht:2009xr}, Higgsless~\cite{Csaki:2003zu,Nomura:2003du,Barbieri:2003pr,Cacciapaglia:2004rb}, gauge-Higgs unification~\cite{Contino:2003ve,Oda:2004rm,Agashe:2004rs,Medina:2007hz,Hosotani:2008tx}, and supergravity~\cite{Zucker:2003qv} models.

A five dimensional gauge theory is not renormalizable and must be seen as an effective field theory. We must take into account all the higher dimensional operators that are allowed by symmetries of a given theory, with appropriate suppression by a cutoff scale~$\Lambda$.
Especially, when there is a bulk scalar field, no symmetry prohibits the existence of the same type of potentials at the orbifold fixed-points as that of the bulk potential (with appropriate rescaling by the cutoff $\Lambda$ to match its mass dimension). To repeat, the five dimensional QFT with a bulk scalar, given as an effective theory, inevitably has the brane potentials.

In~\cite{Haba:2009uu} we stressed the importance of the brane-localized potential and considered an extreme case where the electroweak symmetry breaking is solely due to the brane-localized potential.\footnote{See Refs. \cite{DeCurtis:2002nd,Coradeschi:2007gb} for related works that also take into account the brane-localzed potential. In Ref. \cite{DeCurtis:2002nd}, the equivalence theorem is studied in a two Higgs doublet model with a brane-localized potential. In Ref. \cite{Coradeschi:2007gb}, it has been shown that the vev profile can be non-flat under the presence of a brane-localized potential. In both papers, the KK expansion of the Higgs field is not performed in a diagonal basis and the wave function profile of a KK mass eigenstate was hardly observable.
% The main subject of the current paper is to perform the diagonalization and to present the profile of the eigenmode that differs from the vev profile even for the lowest mode.
} In this paper, we concentrate on the opposite extreme where electroweak symmetry breaking is mainly due to the bulk potential, as in the UED model, and take into account the brane localized potentials as small perturbation.\footnote{In~\cite{Flacke:2008ne}, Flacke, Menon and Phalen have emphasized the importance of the brane-localized interactions in the context of the UED model and especially have analyzed the effect from the existence of the brane-localized kinetic (quadratic) term upon the extra dimensional wave-function profile. The brane-localized potential was written but not taken into account in the calculation of the wave function profile. 
In this paper we continue to concentrate on the effect of the brane localized potential.}
%We stress that 
%the classical and quantum fluctuations obey different equations to determine their wave-function profiles in the extra dimension $S^1/Z_2$. 
%the KK expansion of the Higgs field is performed in a diagonal basis and the wave function profile of a KK mass eigenstate can be obtained. The main subject of the current paper is to perform the diagonalization and to present the profile of the eigenmode that differs from the vev profile even for the lowest mode in the extra dimension $S^1/Z_2$.
One of the main subjects of the current study is to perform  
diagonalization  of eigenmodes in order to present their profiles that  
even leads to a difference between the vev and lowest mode profiles.  
Note that this diagonalization has never been achieved in any kind of  
models, except for our previous study \cite{Haba:2009uu}.

The organization of the paper is as follows.
In the next section, we present our idea by the simplest toy model with a single real scalar field in the bulk, under the presence of the brane-localized potentials.
In Section 3, we compute the Kaluza-Klein (KK) expansions for Higgs fields in the UED model with brane potentials, by taking it as a small perturbation.
In Section 4, we compute the KK expansions for gauge fields similarly.
We show that even though the KK masses are distorted by the brane potential, $\rho$ parameter remains the same as the standard model at the tree level.
In Section 5, we summarize our result and show possible future directions.
In Appendix, we give our gauge fixing procedure and show that extra-dimensional component of the gauge field and the would-be Nambu-Goldstone (NG) modes mix each other because of the position dependent vacuum expectation value (vev) while the four dimensional component of the gauge field does not receive such contribution.

%%%%%%%%%%%%%%%%%%%%%%%%%%%%%%%%%%%%%%%%%%%%%%%%%%%%%%%%%%%%%
\section{VEV and Physical Fields under Brane Potentials}
%%%%%%%%%%%%%%%%%%%%%%%%%%%%%%%%%%%%%%%%%%%%%%%%%%%%%%%%%%%%%
\label{real_scalar}
To clarify our previous observation~\cite{Haba:2009uu}, let us first consider a five dimensional theory with a real bulk scalar field $\Phi$, compactified on a line segment $y\in [0,L]$.\footnote{An orbifold theory on $S^1/Z_2$ can be obtained by identifying its brane-localized potentials with twice the corresponding boundary-localized potentials in the line-segment theory.}
% We write the five dimensional coordinate $X^M=(x^\mu,y)$, with $M,N,\dots$ and $\mu,\nu,\dots$ running for $0,\dots,4$ and $0,\dots,3$, with the extra dimensional coordinate being $y\equiv X^4$. 
The action is given by
\begin{align}
	S	&=	\int d^4x\int_0^L dy\left[
				-{1\over2}(\partial_M\Phi)(\partial^M\Phi)
				-\V(\Phi)
				-\delta(y)\,V_0(\Phi)
				-\delta(y-L)\,V_L(\Phi)\right],
			\label{real_scalar_action}
  \end{align}
where $M,N,\dots$ run for $0,\dots,3;5$, our metric convention is $\eta_{MN}=\diag(-1,1,1,1,1)_{MN}$.
Mass dimensions are $[\Phi]=3/2$, $[\V]=5$, and $[V_0]=[V_L]=4$.\footnote{Note that there can be brane localized kinetic terms too~\cite{Flacke:2008ne} $\propto\delta(y-\eta)(\partial_M\Phi)(\partial^M\Phi)$ with $\eta$ being 0 or $L$, which we neglect for simplicity in this paper.}

The variation of the action is
\begin{align}
	\delta S
		&=	\int d^4x\int_0^L dy\,\delta \Phi\left[
					\Box\Phi
					+\partial_y^2\Phi
					-{\partial \V\over\partial\Phi}
					-\delta(y){\partial V_0\over\partial\Phi}
					-\delta(y-L){\partial V_L\over\partial\Phi}\right]
			+\int d^4x\left[-\delta\Phi\partial_y\Phi\right]_{y=0}^{y=L}
			\nonumber\\
%		&=	\int d^4x\int_0^Ldy\bigg[
%				\delta\Phi_X\left(
%					\Box\Phi_X
%					+\partial_y^2\Phi_X
%					-{\partial \V\over\partial\Phi_X}\right)\nonumber\\
%		&\quad\phantom{\int d^4x\int_0^Ldy\bigg[}
%				+\delta(y)\,\delta\Phi_X\left(\partial_y\Phi_X-{\partial V_0\over\partial\Phi_X}\right)
%				+\delta(y-L)\,\delta\Phi_X\left(-\partial_y\Phi_X-{\partial V_L\over\partial\Phi_X}\right)\bigg],
				\label{S_variation}
  \end{align}
where we have performed the partial integration and we define $\Box\equiv\partial_\mu\partial^\mu=-\partial_0^2+\nabla^2$ with $\mu,\nu,\dots$ running for 0 to 3.
Resultant bulk equation of motion from the variation~\eqref{S_variation} is
\begin{align}
	\Box\Phi+\partial_y^2\Phi-{\partial \V\over\partial\Phi}=0,
		\label{bulk_eom}
  \end{align}
while the boundary condition at $y=0,L$ reads either Dirichlet
\begin{align}
	\left.\delta\Phi\right|_{y=\eta}=0
		\label{Dirichlet}
  \end{align}
or Neumann
\begin{align}
	\left.\left(\mp\partial_y\Phi-{\partial V_\eta\over\partial \Phi}\right)\right|_{y=\eta}=0,
		\label{Neumann}
  \end{align}
where signs above and below are for $\eta=L$ and $0$, respectively, throughout this paper.
%\footnote{For Dirichlet-Neumann condition at $y=0,L$, we extend by $\Phi_{\text{upstairs}}(-y)=-\Phi(y)$ and $\Phi_{\text{upstairs}}(L+y)=\Phi(L-y)$ for $0<y<L$, etc.}
We have four choices of combination of Dirichlet and Neumann boundary conditions at $y=0$ and $L$, namely
\begin{align}
	(D,D),\quad (D,N),\quad (N,D),\quad \text{and}\quad (N,N).
		\label{four_bcs}
  \end{align}
Difference choice of boundary condition corresponds to different choice of the theory. The theory is fixed once one chooses one of the four conditions.

We comment on the relation between the above ``downstairs'' line-segment picture and the orbifold picture.
Sometimes it is convenient to first define fields on a circle $y\in(-L,L]$, or even in the ``upstairs'' picture $y\in(-\infty,\infty)$.
A special Dirichlet condition $\left.\Phi\right|_{y=\eta}=0$ corresponds to the $Z_2$~odd condition $\Phi(x,\eta+y)=-\Phi(x,\eta-y)$ in the orbifolding, while the Neumann condition~\eqref{Neumann} corresponds to the $Z_2$~even one $\Phi(x,\eta+y)=\Phi(x,\eta-y)$ (with the appropriate redefinition of the brane potential by factor two).
The even~($N,N$) and odd~($D,D$) fields in the orbifold picture are given as (see e.g.~\cite{Gherghetta:2000qt})
\begin{align}
	\Phi^{\text{even}}(x,y)
		&=	\Phi(x,|y|),	\\
	\Phi^{\text{odd}}(x,y)
		&=	\epsilon(y)\Phi(x,|y|),
  \end{align}
where $\epsilon(y)=\pm1$ for $\pm y>0$ and $\Phi$ in the r.h.s.\ is the solution to the bulk equation~\eqref{bulk_eom} in $0<y<L$ subject to the boundary conditions~\eqref{Dirichlet} or \eqref{Neumann}.

We utilize the background field method, separating the field into vev and quantum-fluctuation parts:
\begin{align}
	\Phi(x,y)	&=	\Phi^c(x,y)+\phi^q(x,y).
		\label{classical_quantum}
  \end{align}
%The classical field configuration obeys the bulk equation of motion
In order to determine the vev profile, we need to solve the bulk equation of motion
\begin{align}
	\Box\Phi^c+\partial_y^2\Phi^c-{\partial \V\over\partial\Phi}^c
		&=	0,
					\label{classical_bulk_equation}
  \end{align}
with either the Dirichlet boundary condition
\begin{align}
	\left.\delta\Phi^c\right|_{y=\eta}
		&=	0
			\label{classical_Dirichlet_bc}
  \end{align}
or the Neumann boundary condition
\begin{align}
	\left.\left(\mp\partial_y\Phi^c-{\partial V_\eta\over\partial \Phi}^c\right)\right|_{y=\eta}
		&=	0
				\label{classical_Neumann_bc}
  \end{align}
at each brane. Here and hereafter, we utilize the following shorthand notation:
\begin{align}
	{\partial V\over\partial\Phi}^c(x,y)
		&\equiv	\left.{\partial V\over\partial\Phi}\right|_{\Phi=\Phi^c(x,y)}, &
	{\partial^2 V\over\partial\Phi^2}^c(x,y)
		&\equiv	\left.{\partial^2V\over\partial\Phi^2}\right|_{\Phi=\Phi^c(x,y)},
					\label{shorthand_notation}
  \end{align}
etc. 

We put the separation~\eqref{classical_quantum} into the action~\eqref{real_scalar_action} and expand up to the quadratic terms of the %quantum 
 fields~$\phi^q$. 
Note that the Dirichlet boundary condition on the quantum fluctuation reads $\left.\phi^q\right|_{y=\eta}=0$.
After several partial integrations, utilizing the %classical
 equation of motion~\eqref{classical_bulk_equation} with either the Dirichlet $\left.\phi^q\right|_{y=\eta}=0$ or Neumann~\eqref{classical_Neumann_bc} boundary condition, we obtain the free field action up to the quadratic terms in $\phi^q$
\begin{align}
	S_{\text{free}}
		&=	\int d^4x\int_0^Ldy\bigg(
				{1\over2}\phi^q\left[\Box+\partial_y^2-{\partial^2\V\over\partial\Phi^2}^c\right]\phi^q
				\nonumber\\
		&\qquad
				+	{\delta(y)\over2}\phi^q\left[\partial_y-{\partial^2V_0\over\partial\Phi^2}^c\right]\phi^q
				+	{\delta(y-L)\over2}\phi^q\left[-\partial_y-{\partial^2V_L\over\partial\Phi^2}^c\right]\phi^q
					\bigg).
				\label{free_field_action}
  \end{align}
A few comments are in order:
\begin{itemize}
	\item	The free field action~\eqref{free_field_action} is obtained by the expansion up to quadratic orders. Higher order terms $\propto\phi^n$ with $n>2$ are treated as interactions. Kaluza-Klein (KK) expansion will be performed on the free field~$\phi$ with the action~\eqref{free_field_action}. 
	\item	The boundary conditions~\eqref{Dirichlet} and/or \eqref{Neumann} is put on the whole field~\eqref{classical_quantum} when the theory is defined. That is, when the %classical field~
vev $\Phi^c$ obeys Dirichlet condition $\Phi^c=\text{const.}$ at a boundary, the quantum fluctuation also obeys the Dirichilet one $\delta\Phi=\phi^q=0$. When $\Phi^c$ obeys Neumann condition~\eqref{classical_Neumann_bc} at a boundary, the quantum part $\phi^q$ obeys
	\begin{align}
		\left.\left(\mp\partial_y\phi^q
			-{\partial^2V_\eta\over\partial\Phi^2}^c\phi^q
			\right)\right|_{y=\eta}=0,
			\label{bc_for_quantum_fluctuation}
	  \end{align}
	  where above (below) sign is for $y=L$ (0).
	\item The Neumann boundary condition for %classical part~
vev \eqref{classical_Neumann_bc} and for quantum fluctuation~\eqref{bc_for_quantum_fluctuation} are generically different. Therefore in general, \emph{the wave function profile for the %classical
vev and quantum %parts
fluctuation are different from each other}. We will see it more in detail below.
	\item The boundary condition~\eqref{Neumann} on the whole field~\eqref{classical_quantum} contains terms quadratic and higher order in $\phi^q$, such as
	\begin{align}
		{\delta(y)\over2}\left({\partial^3V_0\over\partial\Phi^3}\right)^c(\phi^q)^2.	\label{dropped_bc}
	  \end{align}
These terms are coming from the cubic and higher order brane-localized interactions, which are dropped to obtain the free field action~\eqref{free_field_action}. Note that exactly these terms account for the difference between the boundary conditions for vev and fluctuation. For example, the brane-localized term corresponding to the condition~\eqref{dropped_bc} is
	\begin{align}
		{\delta(y)\over3!}\left({\partial^3V_0\over\partial\Phi^3}\right)^c(\phi^q)^3.
	  \end{align}
	These dropped terms will be treated as boundary-localized interactions that generically mix different KK modes.
  \end{itemize}

Now let us go on to the KK expansion.
On physical ground, we assume that the %classical field configuration 
vev does not depend on the flat four dimensional coordinates~$x^\mu$: $\Phi^c=\Phi^c(y)$. The equation of motion %for the classical fields 
 are then
\begin{align}
	{d^2\Phi^c\over dy^2}(y)-{\partial\V\over\partial\Phi}^c(y)
		&=	0.
		\label{classical_eom}
  \end{align}
Following the Sturm-Liouville theory, we can always expand any function of $y$, subject to one of the four choices of boundary conditions~\eqref{four_bcs}, in terms of the orthonormal basis
\begin{align}
	\phi^q(x,y)
		&=	\sum_n\phi^q_{n}(x)f_{n}(y),
  \end{align}
where $f_{n}(y)$ are eigenfunctions of the Hermitian differential operator in the free action~\eqref{free_field_action}:
\begin{align}
	\left({d^2\over dy^2}-{\partial^2\V\over\partial\Phi^2}^c(y)\right)f_{n}(y)
		&=	-\mu_{n}^2 f_{n}(y).
			\label{KK_equation}
  \end{align}
The eigenvalues $-\mu_{n}^2$ are real but are not necessarily negative at the moment.\footnote{Recall also that they are not degenerate, that is, $-\mu_{n}^2\neq-\mu_{m}^2$ if $n\neq m$.}

For each $n$th mode, there are totally three unknown constants: two integration constants of the second order differential equation~\eqref{KK_equation} and the eigenvalue $-\mu_{n}^2$. Two of the three are fixed by the two boundary conditions at $y=0$ and $L$, while the last one is fixed by the normalization
\begin{align}
	\int_{0}^{L}dy\,f_n(y)f_m(y)	&=	\delta_{nm}.
		\label{downstairs_normalization}
  \end{align}
Consequent mass dimension is $[f_n]=1/2$. % and $[h_n]=1$.
Eventually the free field action~\eqref{free_field_action} is rendered into
\begin{align}
	S_{\text{free}}
		&=	\sum_n\int d^4x\,{1\over2}\phi^q_n(x)\left(\Box-\mu_{n}^2\right)\phi^q_n(x).
  \end{align}

%%%%%%%%%%%%%%%%%%%%%%%%%%%%%%%%%%%%%%%%%%%%%%%
\section{Boundary Potential on Universal Extra Dimension}
%%%%%%%%%%%%%%%%%%%%%%%%%%%%%%%%%%%%%%%%%%%%%%%
\label{UED_section}
In this section, we study the effect of the brane-localized potentials on the UED model~\cite{Appelquist:2000nn,Appelquist:2002wb}.
In the UED model, the KK parity ${L\over2}-y\to {L\over2}+y$ plays a crucial role to make the Lightest KK Particle (LKP) stable so that it can serve as a dark matter candidate.
In this setup, it is convenient to utilize the new coordinate $z\equiv y-{L\over2}$. The KK parity is realized as $z\to-z$.
Hereafter, we rewrite the labels $\eta=L$ and 0, respectively by $+$ and $-$.
The action for the $SU(2)_L$ doublet Higgs field $H$ is now
\begin{align}
	S_H
		&=	\int d^4x\int_{-L/2}^{L/2}dz\left[-(D_MH)^\dagger(D^MH)-\V(H)-\delta(z-{L/2})\,V_+(H)-\delta(z+{L/2})\,V_-(H)\right],
			\label{H_action}
  \end{align}
where $D_M$ is the gauge covariant derivative
\begin{align}
	D_M	&=	\partial_M+ig_5T^aW_M^a+ig'_5YB_M,
  \end{align}
with $Y=1/2$ and $T^a=\sigma^a/2$ on $H$. (As usual, $\sigma^a$ are the Pauli matrices.)
Mass dimensions are $[H]=[W_M^a]=[B_M]=3/2$ and $[g_5]=[g_5']=-1/2$.
In the UED model, extra dimensional components of the gauge fields $W^\pm_5$, $Z_5$ and $A_5$ are odd under orbifold projection, taking $(D,D)$ boundary conditions, while all the other fields are even, taking $(N,N)$ ones.\footnote{In the UED model, $(D,D)$ condition is set such that the fields $W^\pm_5$, $Z_5$ and $A_5$ are vanishing at the boundary. Generically one can consider fixed but non-vanishing value for $(D,D)$ boundary condition. This type of boundary condition for the Higgs field is utilized in~\cite{HOT3}.}

An important point is that, as a non-renormalizable effective field theory in five dimensions, the bulk and brane potentials should contain all the higher dimensional operators, suppressed by a cutoff scale of the five dimensional theory $\Lambda$:
\begin{align}
	\V(H)
		&=	m^2|H|^2+{\lamhat\over\Lambda}{|H|^4}+O(\Lambda^{-4}),
			\label{bulk_pot_expanded}\\
	V_\pm(H)
		&=	m_\pm|H|^2+{\lamhat_\pm\over\Lambda^2}{|H|^4}+O(\Lambda^{-5}),
			\label{brane_pot_expanded}
  \end{align}
where $\lamhat$ and $\lamhat_\pm$ are dimensionless constants.\footnote{\label{bulk_mass_naturalness} Generically one would also expect that $m\sim m_\pm\sim\Lambda$ as an effective theory. Here we do not pursue this so-called ``naturalness problem'' and take $m^2$ and $m_\pm$, being either positive or negative, as free dimensionful parameters.}
(Recall the mass dimensions: $[\V]=5$, $[V_\pm]=4$, and $[H]=3/2$.)
We emphasize that the presence of the brane potential~\eqref{brane_pot_expanded}, which has been overlooked so far, is \emph{inevitable} since no symmetry can prohibit the existence of~\eqref{brane_pot_expanded} when one allows the bulk potential~\eqref{bulk_pot_expanded}.

Note that we have chosen the following basis
\begin{align}
	H
		&=	\begin{pmatrix}\varphi^+\\ \varphi^0\end{pmatrix}
		 = \begin{pmatrix}\varphi^+\\ {1\over\sqrt{2}}\left(\phi+i\chi\right)\end{pmatrix},
\end{align}
in which the real part $\phi$ (of the electrically neutral scalar $\varphi^0$) takes a vev and plays the role of the real scalar $\Phi$ in the previous section.
Using $|H|^2={\phi^2+\chi^2\over2}+\varphi^+\varphi^-$ with $\varphi^-\equiv(\varphi^+)^\dagger$,
let us rewrite the potentials\footnote{In this paper, we neglect all the back-reactions to the background spacetime geometry and shift zero of the potentials freely.}
\begin{align}
	\V	&=	{\lambda\over4}\left(\phi^2+\chi^2+2\varphi^+\varphi^--v_0^2\right)^2+O(\Lambda^{-4}), \label{bulk_potential} \\
	V_\pm
		&=	{\lambda_\pm\over4}\left(\phi^2+\chi^2+2\varphi^+\varphi^--v_\pm^2\right)^2+O(\Lambda^{-5}), \label{brane_potential}
  \end{align}
where we defined
	$\lambda\equiv\lamhat/\Lambda$ and
	$\lambda_\pm\equiv\lamhat_\pm/\Lambda^2$.
The mass dimensions of the new parameters are $[\lambda]=-1$, $[\lambda_\pm]=-2$, and $[v_0^2]=[v_\pm^2]=3$.
Note that the parameters $v_0^2\equiv m^2/\lambda$ and $v_\pm^2\equiv m_\pm/\lambda_\pm$ can be either positive or negative.\footnote{As stated in footnote~\ref{bulk_mass_naturalness}, the bulk mass squared and the brane mass, which can be positive and/or negative, are taken as free dimensionful parameters and hence $v_0^2$ and $v_\pm^2$ are also free parameters.}

In this notation, the vev $\phi^c(z)$ is determined by the bulk equation of motion
\begin{align}
	{d^2\phi^c(z)\over dz^2}
		-\lambda\left({\phi^c(z)}^2-v_0^2\right)\phi^c(z)=0,
		\label{eom_phi}
  \end{align}
with either the Neumann
\begin{align}
	\left.\mp{d\phi^c(z)\over dz}-\lambda_\pm\left({\phi^c(z)}^2-v_\pm^2\right)\phi^c(z)\right|_{z=\pm L/2}=0
		\label{N_phi}
  \end{align}
or Dirichlet
\begin{align}
	\left.\phi^c(z)\right|_{z=\pm L/2}=\text{const.}
		\label{D_phi}
  \end{align}
boundary condition at each end.

Hereafter, we rewrite $h(x,z)\equiv \phi^q(x,z)$ and drop the label ``$^q$'' from other quantum fluctuations:
\begin{align}
	H(x,z)
		&=	\begin{pmatrix}\varphi^+(x,z)\\ {1\over\sqrt{2}}\left[\phi^c(z)+h(x,z)+i\chi(x,z)\right]\end{pmatrix}.
  \end{align}
For reader's ease, we write down the potential quadratic in quantum fluctuation
\begin{align}
	\V(x,z)	&=	{\lambda\over2}\left({\phi^c(z)}^2-v_0^2\right)\left({\chi(x,z)}^2+2\left|\varphi^+(x,z)\right|^2\right)
			+{\lambda\over2}\left(3{\phi^c(z)}^2-v_0^2\right){h(x,z)}^2,  \\
	V_\pm(x)
		&=	\left.{\lambda_\pm\over2}\left({\phi^c(z)}^2-v_0^2\right)\left({\chi(x,z)}^2+2\left|\varphi^+(x,z)\right|^2\right)
			+{\lambda_\pm\over2}\left(3{\phi^c(z)}^2-v_0^2\right){h(x,z)}^2\right|_{z=\pm L/2}.
  \end{align}
Note that linear terms necessarily drop out, due to the equation of motion for the vev (corresponding to Eq.~\eqref{classical_eom}).
The KK expansion for the quantum fluctuations is given as
\begin{align}
	h(x,z)
		&=	\sum_nh_n(x)f^h_n(z), &
	\chi(x,z)
		&=	\sum_n\chi_n(x)f^\chi_n(z), &
	\varphi^+(x,z)
		&=	\sum_n\varphi^+_n(x)f^\varphi_n(z),
  \end{align} 
where $[h_n]=[\chi_n]=[\varphi^+_n]=1$. Here $f_n$ are eigenfunctions of the KK equations
\begin{align}
	\left[{d^2\over dz^2}-\lambda\left(3{\phi^c(z)}^2-v_0^2\right)\right]f^h_n(z)
		&=	-\mu_{h n}^2 f^h_n(z), \label{h_bulk_eq}\\
	\left[{d^2\over dz^2}-\lambda\left({\phi^c(z)}^2-v_0^2\right)\right]f_{n}^X(z)
		&=	-\mu_{Xn}^2 f_{n}^X(z), \label{X_bulk_eq}
  \end{align}
subjecting to the boundary conditions
\begin{align}
	\left.\left[\mp{d\over dz}-\lambda_\pm\left(3{\phi^c(z)}^2-v_\pm^2\right)\right]f^h_n(z)\right|_{z=\pm L/2}=0, \label{h_bc}\\
	\left.\left[\mp{d\over dz}-\lambda_\pm\left({\phi^c(z)}^2-v_\pm^2\right)\right]f_n^X(z)\right|_{z=\pm L/2}=0, \label{X_bc}
  \end{align}
where $X$ stands for the labels $\chi$ and $\varphi$, both giving the same KK expansions in this case without boundary potential.
Results presented in this section correspond to $\xi=1$ in the $R_\xi$ gauge, see Appendix.

%%%%%%%%%%%
\subsection{No brane potential case}
%%%%%%%%%%%
Let us first review the case without any brane potential $V_\pm(H)=0$, as in the original UED model~\cite{Appelquist:2000nn,Appelquist:2002wb}.
In the model, there is only bulk potential~\eqref{bulk_potential}, with $O(\Lambda^{-4})$ terms being neglected.
The solution to the %classical
 equation of motion~\eqref{eom_phi} is
\begin{align}
	\phi^c(z)	&=	v_0. 		\label{obvious_solution}
  \end{align}
Note that obviously $\chi^c(z)=(\varphi^+)^c(z)=0$ is the solution for other modes.
In the original UED model, all the bulk fields are put the $(N,N)$ boundary condition with $V_\pm=0$:
\begin{align}
	\mp{dH\over dz}(\pm L/2)=0,
		\label{bc_wo_boundary_potential}
  \end{align}
which is trivially satisfied by the constant profile~\eqref{obvious_solution}.

The KK equation corresponding to~\eqref{KK_equation} is now 
\begin{align}
	{d^2f^h_n(z)\over dz^2}-2{\lambda v_0^2}f^h_n(z)
		&=	-\mu_{h n}^2f^h_n(z), \\
	{d^2f^X_n(z)\over dz^2}
		&=	-\mu_{Xn}^2f^X_n(z).
  \end{align}  
The $(N,N)$ boundary condition~\eqref{bc_wo_boundary_potential} simply reads
\begin{align}
	\mp {df_n\over dz}(\pm L/2)&=0,
		\label{NN_for_f}
  \end{align}
for all $h$, $\chi$ and $\varphi^\pm$.

There are three possible cases:
\begin{enumerate}
	\item When $\mu_{h n}^2<2\lambda v_0^2$ or $\mu_{Xn}^2<0$, general solutions are
	\begin{align}
		f_n(z)
			&= \alpha_n\cosh(\kappa_nz)+\beta_n\sinh(\kappa_nz),
  	\end{align}
	where $\kappa_n=\sqrt{2\lambda v_0^2-\mu_{h n}^2}$ or $\kappa_n=\sqrt{|\mu_{Xn}^2|}$, respectively. This cannot satisfy the boundary condition~\eqref{NN_for_f}.
	\item When $\mu_{h n}^2=2\lambda v_0^2$ or $\mu_{Xn}^2=0$, general solutions are
	\begin{align}
		f_n(z)
			&= \alpha_n+\beta_nz.
  	\end{align}
	This is conventionally called zero mode and is written with $n=0$. With the boundary condition~\eqref{NN_for_f} and the normalization~\eqref{downstairs_normalization}, we get
\begin{align}
	f_0(z)	&=	\sqrt{1\over L}.
					\label{unperturbed_f0}
  \end{align}
	\item	When $\mu_{h n}^2>2\lambda v_0^2$ or $\mu_{Xn}^2>0$, general solutions with integration constants $\alpha_n,\beta_n$ are
	\begin{align}
		f_n(z)
			&= \alpha_n\cos(k_nz)+\beta_n\sin(k_nz)
			\label{explicit_general_solution}
  	\end{align}
	where $k_n=\sqrt{\mu_{h n}^2-2\lambda v_0^2}$ or $k_n=\mu_{Xn}>0$, respectively.
	With the boundary condition~\eqref{NN_for_f} and the normalization~\eqref{downstairs_normalization}, we obtain
	\begin{align}
		f_n(z)
			&=	\begin{cases}
					\sqrt{2\over L}\cos({k_nz}) & \text{($n$: even),}\\
					\sqrt{2\over L}\sin({k_nz}) & \text{($n$: odd),}
			  	\end{cases}
						\label{unperturbed_fn}
	  \end{align}
	where $k_n=n\pi/L$. The cosine and sine modes are KK parity even and odd, respectively.
\end{enumerate}
To summarize, the Kaluza-Klein mass for $n\geq0$ is given by
\begin{align}
	\mu_{h n}
		&=	\sqrt{k_n^2+2\lambda v_0^2}
		 =	\sqrt{n^2+\left({2\lambda v_0^2\over\mKK}\right)^2}\mKK, 
		 	\label{mu_hn_unperturbed}\\
	\mu_{Xn}
		&=	k_n
		 =	n\mKK,
					\label{mu_Xn_unperturbed}
  \end{align}
where we defined the unit KK mass $\mKK\equiv\pi/L$.

%%%%%%%%%%%
\subsection{Brane Potential as Perturbation: %Classical
VEV Part}
%%%%%%%%%%%
In Ref.~\cite{Haba:2009uu}, we have considered an extreme case where electroweak symmetry breaking is solely due to the brane potential.
Here we concentrate on the opposite limit where brane potential is put as a small perturbation on the above UED model.

Let us start from the bulk potential~\eqref{bulk_potential} and treat the brane potential $V_\pm$ in~\eqref{brane_potential} as a small perturbation of $O(\epsilon)$.
Note that $v_\pm^2$ can be negative here, corresponding to the positive mass term in the brane potential, while $v_0^2$ is always positive by the starting assumption that the symmetry breaking in mainly generated by the bulk potential. We take $v_0>0$ hereafter.
When we are interested solely in the brane mass term, we can take limit $\lambda_\pm\to0$ with fixed $m_\pm=-\lambda_\pm v_\pm^2$.

Firstly the %classical
 equation of motion~\eqref{eom_phi} is not altered. We seek for a solution of the type
\begin{align}
	\phi^c(z)	&=	v_0+\epsilon\phi^c_1(z),
		\label{perturbed_vev}
  \end{align}
where $\phi^c_1(z)$ is a small perturbation and $\epsilon$ is the expansion parameter eventually set to be unity.
%
%Neglecting terms of $O(\epsilon^2)$,
%\begin{align}
%	\epsilon\,{\partial V_\pm\over\partial\Phi_R}
%		&=	\epsilon\,{\lambda_\pm}\left(v_0^2-v_\pm^2\right)v_0, &
%	\epsilon\,{\partial V_\pm\over\partial\Phi_I}
%		 =	0,
%  \end{align}
%\begin{align}
%	\epsilon\,{\partial^2 V_\pm\over\partial\Phi_R^2}
%		&=	\epsilon\,{\lambda_\pm}\left(3v_0^2-v_\pm^2\right), &
%	\epsilon\,{\partial^2V_\pm\over\partial\Phi_R\partial\Phi_I}
%		&=	0, &
%	\epsilon\,{\partial^2 V_\pm\over\partial\Phi_I^2}
%		&=	\epsilon\,{\lambda_\pm}\left(v_0^2-v_\pm^2\right).
%  \end{align}
%
%Expanding the bulk potential up to the linear order in $\Phi^c_1(z)$ and neglecting terms of $O(\epsilon^2)$, we get
%\begin{align}
%	{\partial\V\over\partial\Phi_R}^c
%		&=	\epsilon\,2\lambda v_0^2\Phi_{1R}^c, &
%	{\partial\V\over\partial\Phi_I}^c
%		&=	0,
%  \end{align}
%\begin{align}
%	{\partial^2\V\over\partial\Phi_R^2}^c
%		&=	2\lambda v_0^2+\epsilon\,6\lambda v_0\Phi_{1R}^c, &
%	{\partial^2\V\over\partial\Phi_I^2}^c
%		&=	\epsilon\,2\lambda v_0\Phi_{1R}^c, &
%	\left({\partial^2\V\over\partial\Phi_R\partial\Phi_I}\right)^c
%		&=	0,
%  \end{align}
%where we continue to choose $\Phi_{1I}^c=0$.
%
We put Eq.~\eqref{perturbed_vev} into Eq.~\eqref{eom_phi} to get
\begin{align}
	{d^2\phi^c_{1}\over dz^2}(z)-2\lambda v_0^2\phi^c_{1}(z) &= 0.
		\label{perturbation_eom}
  \end{align}
The general solution is
\begin{align}
	\phi^c_{1}(z)
		&=	A_1\cosh(\kappa z)+B_1\sinh(\kappa z),
			\label{perturbation_y_dep}
  \end{align}
where we define $\kappa\equiv\sqrt{2\lambda}\,v_0$.
Note the mass dimensions $[\kappa]=1$ and $[A_1]=[B_1]=3/2$.
We sometimes trade $\lambda$ by $\kappa$ in the following.

Noting that the brane potential itself is treated as a perturbation of $O(\epsilon)$, the $(N,N)$ boundary condition~\eqref{N_phi} reads:
\begin{align}
			\left.\mp{d\phi^c_{1}(z)\over dz}
			-\lambda_\pm\left(v_0^2-v_\pm^2\right)v_0\right|_{z=\pm L/2}
				&=	0,
					\label{perturbed_bc}
  \end{align}
that is,
\begin{align}
	-\kappa A_1\sinh(\kappa L/2)
	\mp\kappa B_1\cosh(\kappa L/2)
	-\lambda_\pm\left(v_0^2-v_\pm^2\right)
	v_0
		&=	0.
			\label{ABeq}
  \end{align}
When we assume conserved KK parity on our setup, namely $V_+(H)=V_-(H)$ and hence $\lambda_+=\lambda_-$ and $v_+^2=v_-^2$, the solution to Eq.~\eqref{ABeq} simplifies to
\begin{align}
	A_1	&=	-{\lambda_+v_0\left(v_0^2-v_+^2\right)\over\kappa\sinh{\kappa L\over 2}}, &
	B_1	&=	0. 		\label{A1B1}
  \end{align}

To summarize, when the brane potentials respect the KK parity $V_+=V_-$ the vev becomes KK parity even:
\begin{align}
	\phi^c(z)
		&=	v_0+\epsilon\phi_1^c(z)+O(\epsilon^2), &
		\text{with}\quad
	\phi_1^c(z)
		&=	-{\lambda_+v_0\left(v_0^2-v_+^2\right)\over\kappa\sinh{\kappa L\over2}}\cosh(\kappa z).
			\label{phi1c}
%			\nonumber\\
%		&=	v-\epsilon{\lamhat_+(v^2-v_+^2)\over2\sqrt{3\lamhat}\Lambda^{3/2}\sinh{\kappa L\over2}}\cosh(\kappa z)+O(\epsilon^2).
  \end{align}
Recall that $v_\pm^2$ in the perturbation potential $\epsilon V_\pm$ can be negative while we take $v_0>0$ by construction.

%%%%%%%%%%%
\subsection{Brane Potential as Perturbation: Quantum Part}
%%%%%%%%%%%

We treat the brane potential as a perturbation on the eigenvalue problem~\eqref{h_bulk_eq} with the boundary condition~\eqref{h_bc}.
Recall that we are regarding $V_\pm$ as a small perturbation of $O(\epsilon)$:
\begin{align}
	\epsilon V_\pm
		&=	\epsilon{\lambda_\pm\over4}\left[(v_0+h)^2+\chi^2+2\varphi^+\varphi^--v_\pm^2\right]^2+O(\epsilon^2),\\
	\V	&=	{\lambda\over4}\left(2v_0h+h^2+\chi^2+2\varphi^+\varphi^-\right)^2
			+\epsilon \lambda
				\left(2v_0h+h^2+\chi^2+2\varphi^+\varphi^-\right)\left(v_0+h\right)\phi_1^c+O(\epsilon^2).
  \end{align}
We separate the KK wave function of the physical Higgs field into the unperturbed and perturbed parts
\begin{align}
	f^h_n(z)	&=	f_n^{(0)}(z)+\epsilon f_n^{(1)}(z)+O(\epsilon^2),
  \end{align}
where $f_n^{(0)}(z)$ are explicitly given as the r.h.s.\ of Eqs.~\eqref{unperturbed_f0} and \eqref{unperturbed_fn} with the unperturbed eigenvalues $-\mu_n^2$ given by~\eqref{mu_hn_unperturbed}.
Let us write the new perturbed eigenvalues as $-\mu_n^2-\epsilon\Delta_n$, with $\mu_n$ being given by r.h.s.\ of Eq.~\eqref{mu_hn_unperturbed} and $\Delta_n$ being real constant of mass dimension $[\Delta_n]=2$. 
The first order KK equation from Eq.~\eqref{h_bulk_eq} becomes
\begin{align}
	\left({d^2\over dz^2}-2\lambda v_0^2+\mu_n^2\right)f_n^{(1)}(z)
		&=	\left(6\lambda v_0\phi_1^c(z)-\Delta_n\right)f_n^{(0)}(z).
			\label{first_order_KK}
  \end{align}
The boundary condition~\eqref{h_bc} is now, to the first order,
\begin{align}
	\left.\mp{df_n^{(1)}(z)\over dz}\right|_{z=\pm L/2}
		&=	\left.{\lambda_\pm}\left(3v_0^2-v_\pm^2\right)f_n^{(0)}(z)\right|_{z=\pm L/2}.
			\label{first_order_bc}
  \end{align}

%%%%%%%%%%%%%%
\subsubsection{Zero Mode}

Let us first consider the zero mode KK equation from Eq.~\eqref{first_order_KK}
\begin{align}
	{d^2f_0^{(1)}\over dz^2}(z)
		&=	{6\lambda v_0\phi_1^c(z)-\Delta_0\over\sqrt{L}} \nonumber\\
		&=	{6\lambda v_0\left(A_1\cosh(\kappa z)+B_1\sinh(\kappa z)\right)-\Delta_0\over\sqrt{L}},
  \end{align}
where constants $A_1$ and $B_1$ are given by Eq.~\eqref{A1B1} when there is the conserved KK parity.
General solution is
\begin{align}
	f_0^{(1)}(z)
		&=	\alpha_0
			+\beta_0 z
			-{\Delta_0\over2\sqrt{L}}z^2
			+{3\over v_0\sqrt{L}}\left(A_1\cosh(\kappa z)+B_1\sinh(\kappa z)\right),
  \end{align}
where $\alpha_0$ and $\beta_0$ are integration constants of mass dimensions $[\alpha_0]=1/2$ and $[\beta_0]=3/2$, respectively.

Hereafter, we assume the conserved KK parity: $\lambda_+=\lambda_-$ and $v_+^2=v_-^2$, for simplicity. The solution to the boundary condition~\eqref{first_order_bc} is
\begin{align}
	\Delta_0
		&=	{4\lambda_+v_+^2\over L}, &
	\beta_0
		&=	0.
  \end{align}
The zero mode becomes KK parity even. The constant $\alpha_0$ can be fixed by the normalization condition~\eqref{downstairs_normalization}, or to the first order,
\begin{align}
	\int_{-L/2}^{L/2}dz\,f_0^{(0)}(z)f_0^{(1)}(z)=0,
  \end{align}
so that
\begin{align}
	\alpha_0
		&=	{\lambda_+v_+^2\sqrt{L}\over6}
			+{6\lambda_+v_0^2\over\kappa^2L^{3/2}}\left(1-{v_+^2\over v_0^2}\right).
  \end{align}
Recall the mass dimensions $[v_0]=[v_+]=3/2$, $[\kappa]=1$, $[\alpha_0]=1/2$, $[\lambda]=[L]=-1$, and $[\lambda_+]=-2$.
%\begin{align}
%	\alpha_0
%		&=	{\sqrt{L}\over48}\left(\lambda_+(3v^3-v_+^3)+\lambda_-(3v^3-v_-^3)\right)
%			+{\kappa^2L^2-24\over4\kappa L}A_1\sinh{\kappa L\over2}.
%  \end{align}
%\begin{align}
%	\alpha_0
%		&=	{\sqrt{L}\left(-2\lambda_0v_0^3+\lambda_Lv_L^3\right)\over6}
%			+{3\over2\lambda v^3 L^{3/2}}\left(\lambda_0(v^3-v_0^3)+\lambda_L(v^3-v_L^3)\right).
%  \end{align}

%When the brane potential respects the KK parity $V_0=V_L$, we get
%\begin{align}
%	f_0^{(1)}
%		&=	\alpha_0
%			+{\lambda_0v_0^3\over L^{3/2}}\left(\left({L\over2}\right)^2-z^2\right)
%			-{3\lambda_0\left(v^3-v_0^3\right)\over2\sqrt{\lambda v^3L}}
%				{\cosh(\kappa z)\over\sinh(\kappa L/2)}.
%  \end{align}

%%%%%%%%%%%%%%
\subsubsection{Even Modes}
For even $n$, the KK equation~\eqref{first_order_KK} reads
\begin{align}
	\left({d^2\over dz^2}+k_n^2\right)f_n^{(1)}(z)
		&=	\left(6\lambda v_0\phi_1^c(z)-\Delta_n\right)\sqrt{2\over L}\cos (k_nz)\nonumber\\
		&=	-\left(  {3\kappa\lambda_+(v_0^2-v_+^2)\over\sinh{\kappa L\over2}}\cosh(\kappa z)+\Delta_n\right)\sqrt{2\over L}\cos (k_nz).
  \end{align}
Recall $k_n=n\pi/L$. The solution is
\begin{align}
	f_n^{(1)}(z)
		&=	\alpha_n\cos(k_nz)+\beta_n\sin(k_nz)
			-{\Delta_n\over2\sqrt{2L}k_n^2}\left[\cos(k_nz)+2k_nz\sin(k_nz)\right]\nonumber\\
		&\quad
			+{3\sqrt{2}\over v_0\sqrt{L}\left(4k_n^2+\kappa^2\right)}			\left[\kappa^2\phi^c_1(z)\cos(k_nz)+2k_n{\phi^c_1}'(z)\sin(k_nz)\right],
  \end{align}
where $\phi_1^c(z)$ is given in Eq.~\eqref{phi1c}.

The boundary condition~\eqref{first_order_bc} for even $n$ mode is now, to the first order,
\begin{align}
	\left.\mp{df_n^{(1)}(z)\over dz}\right|_{z=\pm L/2}
		&=	{\lambda_+}\left(3v_0^2-v_+^2\right)\sqrt{2\over L}(-1)^{n/2},
  \end{align}
which gives $\beta_n=0$ and
\begin{align}
	\Delta_n
		&=	{8\lambda_+v_+^2\over L}+{24\lambda_+k_n^2(v_0^2-v_+^2)\over L(4k_n^2+\kappa^2)}.
			\label{Delta_n}
  \end{align}
For $n\gg1$, we get $\Delta_n\to{2\lambda_+(3v_0^2+v_+^2)/ L}$.
As in the zero mode case, the constant $\alpha_n$ can be fixed by the normalization condition
\begin{align}
	\alpha_n
		&=	{12\sqrt{2}\lambda_+\left(v_0^2-v_+^2\right)\kappa^2\over L^{3/2}\left(4k_n^2+\kappa^2\right)^2}.
  \end{align}

%%%
\subsubsection{Odd Modes}
%%%
Finally we consider the odd $n$ modes. The KK equation reads
\begin{align}
	\left({d^2\over dz^2}+k_n^2\right)f_n^{(1)}(z)
		&=	\left(6\lambda v_0\phi_1^c(z)-\Delta_n\right)\sqrt{2\over L}\sin(k_nz)\nonumber\\
		&=	-\left(  {3\kappa\lambda_+(v_0^2-v_+^2)\over\sinh{\kappa L\over2}}\cosh(\kappa z)+\Delta_n\right)\sqrt{2\over L}\sin (k_nz)
  \end{align}
and its general solution is
\begin{align}
	f_n^{(1)}(z)
		&=	\alpha_n\cos(k_n z)+\beta_n\sin(k_nz)
			-{\Delta_n\over2\sqrt{2L}k_n^2}\left[\sin(k_nz)-2k_nz\cos(k_nz)\right]\nonumber\\
		&\quad
			+{3\sqrt{2}\over v_0\sqrt{L}\left(4k_n^2+\kappa^2\right)}
			\left[\kappa^2\phi_1^c(z)\sin(k_nz)-2k_n{\phi_1^c}'(z)\cos(k_nz)\right].
  \end{align}
The boundary condition~\eqref{first_order_bc} for odd $n$ mode is now, to the first order,
\begin{align}
	\left.\mp{df_n^{(1)}(z)\over dz}\right|_{z=\pm L/2}
		&=	\pm{\lambda_+}\left(3v_0^2-v_+^2\right)\sqrt{2\over L}(-1)^{(n-1)/2},
  \end{align}
which gives $\alpha_n=0$ and $\Delta_n$ again as in Eq.~\eqref{Delta_n}.
From the normalization, the last constant $\beta_n$ is obtained as 
\begin{align}
	\beta_n
		&=	{12\sqrt{2}\lambda_+\left(v_0^2-v_+^2\right)\kappa^2\over L^{3/2}\left(4k_n^2+\kappa^2\right)^2},
  \end{align}
which is equal to the value of even-mode's $\alpha_n$.

\subsection{KK expansion of physical Higgs}
To summarize, under the presence of small brane-localized potential, the KK expansion is given by
\begin{align}
	f_0^h(z)
		&=	{1\over\sqrt{L}}\left(1+
				{\lambda_+v_+^2L\over6}
				+{6\lambda_+(v_0^2-v_+^2)\over v_0\kappa^2L}
				-{2\lambda_+v_+^2\over L}z^2
				-{3\lambda_+(v_0^2-v_+^2)\over \kappa\sinh{\kappa L\over2}}
					\cosh(\kappa z)
			\right)\\
	f_n^h(z)
		&=	\sqrt{2\over L}\left(
				C_n
				+{3\kappa^2\over v_0(4k_n^2+\kappa^2)}\phi_1^c(z)\right)
			\left\{\begin{matrix}\cos(k_nz)\\ \sin(k_nz)\end{matrix}\right\}\nonumber\\
		&\quad
			+\left(
				{\Delta_n\over\sqrt{2L}k_n}z
				-\sqrt{\frac{2}{L}}{6k_n\over v_0(4k_n^2+\kappa^2)}{\phi_1^c}'(z)\right)
			\left\{\begin{matrix}\left(-\sin(k_nz)\right)\\ \cos(k_nz)\end{matrix}\right\}
		\qquad \text{for }\begin{cases} \text{$n$: even positive,} \\\text{$n$: odd,}\end{cases}
  \end{align}
where $\epsilon=1$, and $\phi_1^c(z)$ and $\Delta_n$ for $n>0$ are given in 
Eqs.~\eqref{phi1c} and \eqref{Delta_n}, respectively, and 
\begin{align}
	C_n
		&=	1
			+{12\lambda_+\left(v_0^2-v_+^2\right)\kappa^2\over L\left(4k_n^2+\kappa^2\right)^2}
			-{\Delta_n\over4k_n^2}.
  \end{align}
The perturbed KK mass becomes, respectively for $n=0$ and $n>0$,
\begin{align}
	\mu_{h0}^2
		 =	\kappa^2+\Delta_0
		&=	2\lambda v_0^2+{4\lambda_+v_+^2\over L},\\
	\mu_{hn}^2
		 =	k_n^2+\kappa^2+\Delta_n
		&=	\left({\pi n\over L}\right)^2+2\lambda v_0^2+\Delta_n.
  \end{align}

The case where we have only positive mass term on the brane $V_+=m_+|H|^2={m_+\over2}\phi^2+\cdots$ can be obtained by taking limit $\lambda_+\to0$ with fixed $m_+=-\lambda_+v_+^2>0$:
\begin{align}
	\phi_1^c(z)
		&\to	-{m_+v_0\over\kappa\sinh{\kappa L\over 2}}\cosh(\kappa z),\\
	\Delta_0
		&\to	-{4m_+\over L},\\
	\Delta_{n>0}
		&\to	-{8m_+\over L}{k_n^2+\kappa^2\over4k_n^2+\kappa^2},\\
	C_n
		&\to	1+{12m_+\kappa^2\over L(4k_n^2+\kappa^2)^2}+{2m_+(k_n^2+\kappa^2)\over  L k_n^2(4k_n^2+\kappa^2)}.
  \end{align}
For a very high KK mode $n\gg1$, the limit further simplifies to
\begin{align}
	\Delta_{n>0}
		&\to	-{2m_+\over L}, &
	C_n
		&\to	1.
  \end{align}

%%%%%%%%%%%%%%%%%%%%%%%%%%%%%%%%%%%%%%%%%%%%%%%%%%%%%%
\section{Bulk Gauge Field under Higgs Brane Potential}
%%%%%%%%%%%%%%%%%%%%%%%%%%%%%%%%%%%%%%%%%%%%%%%%%%%%%%
Under the presence of the brane potential, the vev of the Higgs field is distorted as in Eq.~\eqref{phi1c} so that it has non-trivial extra dimensional profile.
Let us see how the gauge field wave function is modified in this case.

As shown in Appendix, the position dependent vev $v(z)\equiv\phi^c(z)$ generates the position dependent bulk mass terms for the gauge fields $W^\pm_\mu$ and $Z_\mu$.
When KK-expanding as
\begin{align}
	W^\pm_\mu(x,z)
		&=	\sum_nf^W_n(z)W^\pm_{n\mu}(x), &
	Z_\mu(x,z)
		&=	\sum_nf^Z_n(z)Z_{n\mu}(x),	
  \end{align}
resultant bulk KK equation becomes
\begin{align}
	\left({d^2\over dz^2}-m_V^2(z)\right)
		f^V(z) &=	-\mu_{Vn}^2f^V(z),
  \end{align}
where the label $V$ stands for $W$ and $Z$.
In contrast, their boundary conditions are not modified from the ordinary $(N,N)$ ones
\begin{align}
	\left.{df^V(z)\over dz}\right|_{z=\pm L/2}
		&=	0,
			\label{WZ_bc}
  \end{align}
since we neglect the brane-localized Higgs kinetic terms in our analysis.

Again let us solve the KK equation iteratively by taking the Higgs brane potential as small perturbation.
From Eq.~\eqref{phi1c}, we see
\begin{align}
	m_V^2	&=	m_{V0}^2+\epsilon g_Vm_{V0}\phi_1^c(z)+O(\epsilon^2), &
  \end{align}
where we define
\begin{align}
	m_{W0}	&\equiv	{gv_0\over2}, &
	m_{Z0}	&\equiv	{g_Zv_0\over2},
  \end{align}
with $g_Z\equiv\sqrt{g^2+{g'}^2}$.
The zeroth order solution with the boundary condition~\eqref{WZ_bc} is, both for $W$ and $Z$,
\begin{align}
	f^{(0)}_0(z)
		&=	{1\over\sqrt{L}}, \\
	f^{(0)}_n(z)
		&=	\begin{cases}
				\sqrt{2\over L}\cos(k_nz) & \text{for $n$: even positive},\\
				\sqrt{2\over L}\sin(k_nz) & \text{for $n$: odd},
			  \end{cases}
  \end{align}
where again $k_n=\pi n/L$ and the zeroth order KK masses are given by
\begin{align}
	\mu_{Vn}^2	&=	k_n^2+m_{V0}^2.
  \end{align}

Writing the eigenvalues of the KK equation $-\mu_{V0}^2-\epsilon\Delta^V_n$, the first order KK equation for the eigenfunction $f^{(0)}_n(z)+\epsilon f^{V(1)}_n(z)$ is
\begin{align}
	\left({d^2\over dz^2}+k_n^2\right)f^{V(1)}_n(z)
		&=	\left(g_Vm_{V0}\phi_1^c(z)-\Delta^V_n\right)f^{(0)}_n(z).
  \end{align}
The solution subjecting to the boundary condition~\eqref{WZ_bc} is obtained similarly to the Higgs case
\begin{align}
	f_0^{V(1)}(z)
		&=	{1\over\sqrt{L}}\left(
				\alpha_0^V
				-{\Delta_0^V\over2}z^2
				-{2m_V^2\lambda_+(v_0^2-v_+^2)\over\kappa^3\sinh{\kappa L\over 2}}\cosh(\kappa z)
				\right), \\
	f_{n>0}^{V(1)}(z)
		&=	\sqrt{2\over L}\left(
				\alpha^V_n
				-{\Delta^V_n\over4k_n^2}
				+{g_Vm_{V0}\over4k_n^2+\kappa^2}\phi_1^c(z)
				\right)
			\left\{\begin{matrix}\cos(k_nz)\\ \sin(k_nz)\end{matrix}\right\}\nonumber\\
		&\quad
			+\sqrt{2\over L}\left(
				{\Delta^V_n\over2k_n}z
				-{2g_Vm_{V0}k_n\over(4k_n^2+\kappa^2)\kappa^2}{\phi_1^c}'(z)
				\right)
			\left\{\begin{matrix}\left(-\sin(k_nz)\right)\\ \cos(k_nz)\end{matrix}\right\},
  \end{align}
where
\begin{align}
	\Delta_0^V
		&=	-{4m_{V0}^2\lambda_+(v_0^2-v_+^2)\over \kappa^2L}, \\
	\Delta_n^V
		&=	-{8m_{V0}^2\lambda_+(v_0^2-v_+^2)(2k_n^2+\kappa^2)\over L\kappa^2(4k_n^2+\kappa^2)},
  \end{align}
and
\begin{align}
	\alpha^V_0
		&=	{m_{V0}^2\lambda_+(v_0^2-v_+^2)(24-\kappa^2L^2)\over6\kappa^4L},\\
	\alpha^V_n
		&=	{8m_{V0}^2\lambda_+\left(v_0^2-v_+^2\right)\over L\left(4k_n^2+\kappa^2\right)^2}.
  \end{align}

When there is only positive mass term on the brane $V_+=m_+|H|^2$, the solution is obtained by taking limit $\lambda_+\to0$ with fixed $m_+\equiv-\lambda_+v_+^2$
\begin{align}
	\Delta_0^V
		&\to	-{4m_{V0}^2m_+\over \kappa^2L}, \\
	\Delta_n^V
		&\to	-{8m_{V0}^2m_+(2k_n^2+\kappa^2)\over L\kappa^2(4k_n^2+\kappa^2)},
  \end{align}
and
\begin{align}
	\alpha^V_0
		&=	{m_{V0}^2m_+(24-\kappa^2L^2)\over6\kappa^4L},\\
	\alpha^V_n
		&=	{4m_{V0}^2m_+\over L\left(4k_n^2+\kappa^2\right)^2}.
  \end{align}
For a very high KK modes $n\gg1$, they further simplify to
\begin{align}
	\Delta_n^V
		&\to	-{4m_{V0}^2m_+\over L\kappa^2},\\
	\alpha^V_n
		&\to	0.
  \end{align}

We note that the observed physical mass-squared for $W^\pm$ and $Z$ bosons correspond to $m_{V0}^2+\Delta_0^V$.
Since the correction to the gauge boson mass-squared $\Delta_0^V$ is proportional to $m_{V0}^2$, the correction to the $W$ and $Z$ masses are proportial to the corresponding gauge coupling $g$ and $g_Z$, respectively, with the uniform coefficient $-{4v_0^2\lambda_+(v_0^2-v_+^2)\over\kappa^2L}$.
Therefore, the ratio of the $W$ and $Z$ boson masses are still proportional to the ratio of the gauge coupling $g/g_Z$.
The brane localized Higgs potential does not change the $\rho$ parameter of the model even though it does change the mass formula,
as is expected from the fact that the introduction of the brane potential does not violate the custodial symmetry. 

%Finally, let us comment that our UED model with the brane localized potential does not suffer from serious FCNC problem since the model has only one Higgs doublet. It might be worth studying this issue further in our model, along with the line of Refs. \cite{Buras:2003wc,Mohanta:2006ae,Colangelo:2007jy}.

%Finally, let us comment that our UED model with the brane localized
%potential does not suffer from serious FCNC problem since the model
%has only one Higgs doublet. We note that in our model, all the bulk 
%fermions have a flat wave function profile and that there are no 
%tree level FCNC processes coming from the overlap integral along 
%the extra dimension. This is not affected by the fact that Higgs 
%vev and physical field have different profile. One might think that 
%contribution from a loop of charged KK Higgs field is different 
%from that in the UED model. However, the only difference is coming 
%from the change of the boundary condition from Neumann to Dirichlet.
%Therefore, the overlap integral of two $n$th modes and a single (flat)
%zero mode are identical in these two models, giving the same coupling.
%FCNC loop contributions from KK modes are not different from the
%ordinary UED model \cite{Buras:2003wc,Mohanta:2006ae,Colangelo:2007jy}.

%%%%%%%%%%%%%%%%%%%%%%%%%%%%%%%%%
\section{Summary and Discussions}
%%%%%%%%%%%%%%%%%%%%%%%%%%%%%%%%%
We have further clarified our previous observation that the brane localized potential can make the extra-dimensional profiles of the vev and lowest KK mode different from each other.
%\textcolor{red}{The KK expansion of the Higgs field has been performed in a diagonal basis and the wave function profile of a KK mass eigenstate could be obtained.}
One of the main subjects of this paper is to perform diagonalization  
of eigenmodes in order to present their profiles that even leads to a  
difference between the vev and lowest mode profiles. We note that this  
diagonalization has never been achieved in any kind of models, except  
for our previous study \cite{Haba:2009uu}. Especially we have explained what makes the difference from the view point of free part of the Lagrangian.

We have considered the UED model and obtained the KK expansion for the Higgs field, under the presence of the brane-localized potential.
We find that small boundary potential raises the KK masses when it is wine-bottle shape with negative mass-squared at its origin, while it lowers the KK masses when there is only positive mass term.
KK parity is conserved in all the modes by introduction of the KK parity even potential $V_+=V_-$.

We have also computed the KK expansion for the four dimensional components of the gauge fields $W_\mu^\pm$ and $Z_\mu$.
Contrary to the Higgs field case, gauge boson KK masses acquire negative contribution for both the wine-bottle and positive-mass shapes of boundary potential.
Even though $W_\mu^\pm$ and $Z_\mu$ have different position-dependent bulk masses and hence the oscillation of their wave function is different in the extra dimension, the resultant $\rho$ parameter remains the same.
This reflects the fact that the custodial symmetry remains intact under the presence of the boundary potential.

It would be interesting to compute the KK expansions of extra dimensional component of gauge fields and the would-be NG modes as well as the bulk fermions, whose masses are modified by the position dependent vev too.
It is also worth studying the brane-localized Higgs kinetic term simultaneously in our setup.
These subjects will be treated in a separate publication.

\subsection*{Acknowledgement} 
We would like to thank T. Yamashita for very helpful discussions,
and also thank S. Matsumoto for useful discussions.
This work is partially supported by Scientific Grant by Ministry of
Education and Science, Nos.\ 20540272, 20039006, 20025004,
20244028, and 19740171. The work of RT is supported by
 the GCOE Program, The Next Generation of Physics, Spun from Universality and 
Emergence.

\appendix
\section*{Appendix}

%%%%%%%%%%%%%
\section*{Gauge Fixing}

Basically we follow the notation of Ref.~\cite{Haba:2009uu}, summarized in its Appendix C, except for the normalization of the vev $v$ which differs by a factor $\sqrt{2}$.
In our basis
\begin{align}
	H^c	&=	\begin{pmatrix}0\\ {v(z)\over\sqrt{2}}\end{pmatrix}, &
	H^q	&=	\begin{pmatrix}\varphi^+\\ {h(x,z)+i\chi(x,z)\over\sqrt{2}}\end{pmatrix}, &
  \end{align}
where we have rewritten the vev $v(z)\equiv \phi^c(z)$.
The covariant derivative on the Higgs field is
\begin{align}
	D_MH
		&=	\partial_MH
			+{ig\over\sqrt{2}}
				\begin{pmatrix}
				0 & W^+_M \\
				W^-_M & 0
				\end{pmatrix}H
			+ie\begin{pmatrix}
				{1\over\tan2\theta_W}Z_M+A_M & 0 \\
				0 & -{1\over\sin2\theta_W}Z_M
				\end{pmatrix}H\nonumber\\
		&=	\begin{pmatrix}
				\partial_M\varphi^+\\
				{\partial_Mv+\partial_Mh+i\partial_M\chi\over\sqrt{2}}
			  \end{pmatrix}
			+{ig\over\sqrt{2}}\begin{pmatrix}
				W_M^+{v+h+i\chi\over\sqrt{2}}\\
				W_M^-\varphi^+
			   \end{pmatrix}
			+ie\begin{pmatrix}
				\left({1\over\tan2\theta_W}Z_M+A_M\right)\varphi^+\\
				-{1\over\sin2\theta_W}Z_M{v+h+i\chi\over\sqrt{2}}
				\end{pmatrix},
  \end{align}
where we have defined
\begin{align}
	W^\pm_M
		&=	{W^1_M\mp iW^2_M\over\sqrt{2}}, &
	\begin{pmatrix}Z_M\\ A_M\end{pmatrix}
		&=	\begin{pmatrix}c & -s \\ s & c\end{pmatrix}\begin{pmatrix}W^3_M\\ B_M\end{pmatrix},
  \end{align}
with
\begin{align}
	c	&\equiv	\cos\theta_W
		=			{g\over\sqrt{g^2+{g'}^2}}, &
	s	&\equiv	\sin\theta_W
		=			{g'\over\sqrt{g^2+{g'}^2}}, &
	e	&\equiv	{gg'\over\sqrt{g^2+{g'}^2}}.
  \end{align}
Note that the bulk gauge boson masses $m_W$ and $m_Z$ are $z$~dependent now
\begin{align}
	m_W(z)
		&\equiv	{gv(z)\over2},	&
	m_Z(z)
		&\equiv	{\sqrt{g^2+{g'}^2}\over2}v(z)
		=			{e\over\sin2\theta_W}v(z).
  \end{align}
Mass dimensions are $[g]=[g']=[e]=-1/2$ and $[v]=[W^\pm_M]=[Z_M]=[A_M]=3/2$.
The Higgs kinetic Lagrangian is
\begin{align}
	\mathcal{L}_H
		 =	-\left|D_MH\right|^2
		&=	-\left|
		 		\partial_M\varphi^+
				+im_WW_M^+
				+{ig\over2}W_M^+\left(h+i\chi\right)
				+ie\left({1\over\tan2\theta_W}Z_M+A_M\right)\varphi^+
				\right|^2\nonumber\\
		&\quad
			-{1\over2}\left|
				\partial_Mv+\partial_Mh+i\partial_M\chi
				+igW_M^-\varphi^+
				-im_ZZ_M
				-{ie\over\sin2\theta_W}Z_M\left(h+i\chi\right)
			\right|^2,
  \end{align}
where the contraction of the Lorentz indices is understood. 
The quadratic terms are
\begin{align}
	\mathcal{L}_H^{\text{quad}}
		&=	-\left|\partial_M\varphi^+\right|^2
			-{\left(\partial_Mh\right)^2+\left(\partial_M\chi\right)^2\over2}
			-m_W^2\left|W_M^+\right|^2
			-{m_Z^2\over2}(Z_M)^2\nonumber\\
		&\quad
			+im_W\left(
				W^{-M}\partial_M\varphi^+
				-W^{+M}\partial_M\varphi^-
				\right)
			+m_ZZ^M\partial_M\chi\nonumber\\
		&\quad
			-\left(\partial_5v\right)\left(
				\partial_5h
				+{ig\over2}\left(W_5^-\varphi^+-W_5^+\varphi^-\right)
				+{e\over\sin2\theta_W}Z_5\chi\right).
  \end{align} 
The terms in the last line are coming from the non-trivial profile of the vev in the extra dimension. 
  
We employ the following $R_\xi$-like gauge fixing\footnote{When we also introduce brane-localized Higgs kinetic terms, we need to add extra gauge fixing terms localized on the branes.}
\begin{align}
	\mathcal{L}_{\text{GF}}
		&=	-{1\over2\xi}\left(\sum_{a=1}^3f^af^a+f^Bf^B\right),
  \end{align}
where
\begin{align}
	f^a
		&=	\partial_M W^{aM}
			+ig\xi\left({H^q}^\dagger T^aH^c-{H^c}^\dagger T^aH^q\right),\nonumber\\
	f^B
		&=	\partial_M B^M
			+ig'\xi\left({H^q}^\dagger YH^c-{H^c}^\dagger YH^q\right).
  \end{align}
By the redefinition
\begin{align}
	f^\pm	\equiv	{f^1\mp if^2\over\sqrt{2}}
			&=	\partial_M W^{\pm M}
			 	\mp i\xi m_W\varphi^\pm, \\
	f^Z	\equiv	cf^3-sf^B
			&=	\partial_M Z^M
				-\xi m_Z\chi,\\
	f^A	\equiv	sf^3+cf^B
			&=	\partial_M A^M,
  \end{align}
we can rewrite
\begin{align}
	\mathcal{L}_{\text{GF}}
		&=	-{1\over\xi}f^+f^-
		 	-{1\over2\xi}\left(f^Zf^Z+f^Af^A\right) \nonumber\\
		&=	-{1\over\xi}\left|\partial_MW^{+M}\right|^2
			-{1\over2\xi}\left(
				\left(\partial_MZ^M\right)^2
				+\left(\partial_MA^M\right)^2
				\right)\nonumber\\
		&\quad
			+im_W\left(\varphi^+\partial_MW^{-M}-\varphi^-\partial_MW^{+M}\right)
			+m_Z\chi\partial_MZ^M
				-\xi m_W^2\left|\varphi^+\right|^2
				-{\xi m_Z^2\over 2}\chi^2.
  \end{align}
The following gauge choices can be considered.
\begin{enumerate}
\item
For $\xi=1$, the sum of quadratic terms simplifies to
\begin{align}
	\mathcal{L}_{H+\text{GF}_{\xi=1}}^{\text{quad}}
		&=	-\left|\partial_M\varphi^+\right|^2
			-m_W^2\left|\varphi^+\right|^2
			-{1\over2}\left(\partial_M\chi\right)^2
			-{m_Z^2\over2}\chi^2
			-{1\over2}\left(\partial_Mh\right)^2\nonumber\\
		&\quad
			-\left|\partial_NW^{+N}\right|^2
			-m_W^2\left|W_M^+\right|^2
			-{1\over2}\left(\partial_NZ^N\right)^2
			-{m_Z^2\over2}(Z_M)^2
			-{1\over2}\left(\partial_MA^M\right)^2\nonumber\\
		&\quad
			+\partial_5\left[im_W\left(W^-_5\varphi^+-W^+_5\varphi^-\right)+m_ZZ_5\chi\right]
			\nonumber\\			
		&\quad
			-\left(\partial_5v\right)\left(\partial_5h\right)
			-2\left(\partial_5v\right)\left(
				{ig\over2}\left(W_5^-\varphi^+-W_5^+\varphi^-\right)
				+{e\over\sin2\theta_W}Z_5\chi\right).
				\label{quad_terms}
  \end{align}
The third (second last) line is a total derivative and potentially contributes as boundary localized mixing terms between gauge fields and the would-be NG modes when we integrate out the extra dimension for the KK reduction where the vev is independent of four-dimensional spacetime coordinate. 
In the UED model of our current consideration, all the extra dimensional components of a vector field are assumed to be odd under the orbifold projection $y\to -y$ and take the following Dirichlet boundary conditions
\begin{align}
	\left.W_5^\pm\right|_{z=\pm L/2}
		=	\left.Z_5\right|_{z=\pm L/2}
		=	\left.A_5\right|_{z=\pm L/2}
		=	0.
  \end{align}
Under this assumption, the third line can be safely neglected.

The last line in Eq.~\eqref{quad_terms} is due to the non-trivial wave function profile of the vev, which mixes the extra-dimensional component of the gauge fields and the would-be NG modes.
The first term in the last line $-(\partial_5v)(\partial_5h)$ is treated properly in Secs.~\ref{real_scalar} and \ref{UED_section}, while impact from the other mixing terms will be presented elsewhere.

\item
In the unitary gauge $\xi\to\infty$, the would-be NG bosons $\varphi^\pm$ and $\chi$ become infinitely heavy and decouple
\begin{align}
	\mathcal{L}_{H+\text{GF}}^{\text{quad}}
		&\to
			-{1\over2}\left(\partial_Mh\right)^2
			-m_W^2\left|W_M^+\right|^2
			-{m_Z^2\over2}(Z_M)^2.
				\label{quad_terms_in_unitary_gauge}
  \end{align}
\end{enumerate}
Hereafter, we employ the $\xi=1$ gauge.

The gauge kinetic Lagrangian is
\begin{align}
	\mathcal{L}_{\text{YM}}
		&=	-{1\over4}\left(\sum_{a=1}^3F^a_{MN}F^{aMN}+F^B_{MN}F^{B\,MN}\right).
  \end{align}
From the redefinition
\begin{align}
	W^\pm_M
		&=	{W^1_M\mp iW^2_M\over\sqrt{2}}, \\
	Z_M
		&=	cW^3_M-sB_M, \\
	A_M
		&=	sW^3_M+cB_M,
  \end{align}
we get
\begin{align}
	F^\pm_{MN}
		&=	\partial_MW^\pm_N-\partial_NW^\pm_M
			\pm 2ig\left(W_M^3W^\pm_N-W_N^3W_M^\pm\right),\\
	F^3_{MN}
		&=	\partial_MW^3_N-\partial_NW^3_M
			+2g\left(W^+_MW^-_N-W^+_NW^-_M\right),
  \end{align}
with $W^3_M=cZ_M+sA_M$, and
\begin{align}
	\mathcal{L}_{\text{YM}}
		&=	-{1\over2}F^+_{MN}F^{-MN}
			-{1\over4}\left[
				F^3_{MN}F^{3\,MN}+F^B_{MN}F^{B\,MN}\right].
  \end{align}
Quadratic terms are\footnote{We do not consider Wilson-line phases and put all the vevs of gauge field zero.}
\begin{align}
	\mathcal{L}_{\text{YM}}^{\text{quad}}
		&=	-{1\over2}\sum_{a=1}^3\left(
				-W^a_\mu\Box W^{a\mu}
				-\left(\partial_\mu W^{a\mu}\right)^2
				+\left(\partial_5W^a_\mu\right)\left(\partial_5W^{a\mu}
					\right)
				-W_5^a\Box W_5^a
				+2W_5^a\partial_5\left(\partial_\mu W^{a\mu}\right)
				\right)\nonumber\\
		&\quad
			-{1\over2}\left(
				-B_\mu\Box B^{\mu}
				-\left(\partial_\mu B^{\mu}\right)^2
				+\left(\partial_5B_\mu\right)\left(\partial_5B^{\mu}\right)
				-B_5\Box B_5
				+2B_5\partial_5\left(\partial_\mu B^{\mu}\right)
				\right)\nonumber\\
		&=	-\left[
				-W^+_\mu\Box W^{-\mu}
				-\left|\partial_\mu W^{+\mu}\right|^2
				+\left(\partial_5 W^+_\mu\right)\left(\partial_5 W^{-\mu}\right)
				\right] \nonumber\\
		&\quad
			-\left[
				-W_5^+\Box W_5^-
				+W_5^+\partial_5\left(\partial_\mu W^{-\mu}\right)
				+W_5^-\partial_5\left(\partial_\mu W^{+\mu}\right)
				\right]\nonumber\\
		&\quad
			-{1\over2}\left(
				-Z_\mu\Box Z^{\mu}
				-\left(\partial_\mu Z^{\mu}\right)^2
				+\left(\partial_5Z_\mu\right)\left(\partial_5Z^{\mu}\right)
				-Z_5\Box Z_5
				+2Z_5\partial_5\left(\partial_\mu Z^{\mu}\right)
				\right)\nonumber\\
		&\quad
			-{1\over2}\left(
				-A_\mu\Box A^{\mu}
				-\left(\partial_\mu A^{\mu}\right)^2
				+\left(\partial_5A_\mu\right)\left(\partial_5A^{\mu}\right)
				-A_5\Box A_5
				+2A_5\partial_5\left(\partial_\mu A^{\mu}\right)
				\right).
  \end{align}

\end{document}